\documentclass[journal]{IEEEtran}
\usepackage{mathrsfs}
\usepackage{amsfonts}
\usepackage{ifpdf}
\usepackage{cite}
\ifCLASSINFOpdf
\usepackage[pdftex]{graphicx}
\else \fi
\usepackage[cmex10]{amsmath}
\usepackage{array}
\usepackage{mdwmath}
\usepackage{mdwtab}
\usepackage{eqparbox}
\usepackage[tight,footnotesize]{subfigure}
\usepackage{amsmath}
\usepackage{epsfig}
\usepackage{ae}
\usepackage{amssymb}
\usepackage{times}
\usepackage{amsmath}   % add
\usepackage{booktabs}
\usepackage{color}
\usepackage{bm}
\usepackage{mathtools}
\usepackage{epstopdf}
\usepackage{subeqnarray}
\usepackage{cases}

\usepackage{algorithmic}
\usepackage{algorithm}
\usepackage{threeparttable}
\usepackage{multicol}
\usepackage{verbatim}
\usepackage{stfloats}
\usepackage{epstopdf}

\begin{document}
\title{IS2N: Intent-Driven Security Software-Defined Network with Blockchain}
\author{Yanbo Song\authorrefmark{1}, Tao Feng\authorrefmark{2},
Chungang Yang\authorrefmark{1}, Xinru Mi\authorrefmark{1}, Shanqing
Jiang\authorrefmark{3}, Mohsen Guizani\authorrefmark{4}
\thanks{\authorrefmark{1} Y. Song, C. Yang, and X. Mi are with the State Key Laboratory on Integrated Services
Networks, Xidian University, Xi'an, 710071 China (emails: songyanbo94@163.com;
chgyang2010@163.com; xinrum@163.com;)}
\thanks{\authorrefmark{2} F. Tao is with Institute of System Engineering, Academy of Military Sciences PLA, Beijing, China
(email: feng09@163.com).}  
\thanks{\authorrefmark{3} Q. Jiang is with School of Cyber Science and Engineering, Southeast University, Nanjing, China (email: sqjiang@njnet.edu.cn).}
%}
\thanks{\authorrefmark{4} M. Guizani is with Mohamed Bin Zayed University of Artificial Intelligence (MBZUAI), UAE (email: mguizani@ieee.org).}

\thanks{This work was supported in part by the National Key Research and Development Program of China (2020YFB1807700).}

} \maketitle

\begin{abstract}
	
Software-defined network (SDN) is characterized by its programmability, flexibility, and the separation of control and data planes. However, SDN still have many challenges, particularly concerning the security of network information synchronization and network element registration. Blockchain and intent-driven networks are recent  technologies to establish secure and intelligent SDN. This article investigates the blockchain-based architecture and intent-driven mechanisms to implement intent-driven security software-defined networks (IS2N). Specifically, we propose a novel four-layer architecture of the IS2N with security capabilities. We integrate an intent-driven security management mechanism in the IS2N to achieve automate network security management. Finally, we develop an IS2N platform with blockchain middle-layer to achieve security capabilities and security store network-level snapshots, such as device registration and OpenFlow messages. Our simulations show that IS2N is more flexible than conventional strategies at resolving problems during network operations and has a minimal effect on the SDN.

\end{abstract}


\section{Introduction}

Software-defined network (SDN) is a component of the evolution of the future network. The network controller, a software-based entity with logical centralization, receives all information and control logic from network devices and is located in the control plane. The programmability and flexibility of SDN may improve the network's security, but it also has the potential to cause new security breaches\cite{1}.

The SDN has two different kinds of control modes: in-band and out-of-band. The data and control paths share the same bandwidth in in-band mode, which does not connect all switches directly to the controller. Since security attacks on the data plane mainly influence the network transmission performance and the quality of service assurance, attacks on the network control system have a direct impact on its operation, even resulting in paralysis. If the controller is hijacked, it loses the ability to operate the network, particularly in the in-band mode. And even though the controller can change the data path by forwarding the Flow-Mod message to the switches when the data path is disrupted. Therefore, security of the controller and the control plane is vital. Verifying a controller's identity is the first line of defense to ensure their access to the network securely. 

As the network scale grows, the SDN transforms from conventional single-controller and multi-controller patterns to cluster and distribution ones that manage more devices. However, if the network is divided into multiple software-defined sub-networks, network-level information in the controllers, for example, topology and link information, will be challenging to synchronize and maintain among them. Controlled based on inconsistent data is meaningless. Moreover, a control channel could be a possible option for attacking the network and directly impacting the data plane. Vulnerable network protocols can poison the topology view, which can facilitate the execution of attacks on the data plane. Further, attackers may stealthily manipulate traffic by OpenFlow rules, leading to active network attacks such as a man-in-the-middle attack. Static connections between switches and controllers lack a security defense mechanism. Attackers can exploit weak or nonexistent access control mechanisms to gain unsupervised access to the SDN elements, including unauthorized access, disclosure of sensitive information, and network modification. Therefore, implementing a security authentication mechanism is crucial in the security field.

In general, the novel security challenges in the SDN may be divided into three main categories as follows:

 \begin{itemize}
    \item How to safeguard the consistency of the network information?
    \item How to introduce the security to the control plan?
    \item How to maintain the trust in network elements?
 \end{itemize}

In recent times, network security is an important foundational characteristic. This development is attributable to emerging technologies such as Blockchain and Intent-Driven Networks (IDNs) which have ushered a paradigm shift in conventional communication networks. In particular, the security of SDN has been enhanced through the incorporation of Blockchain\cite{azab2019towards}. Previous research has demonstrated the application of Blockchain as a certification authority to illustrate the advantages of the technology \cite{4}. Moreover, the IDN concept advances network management automation by abstracting network strategy, a notion espoused by Gartner \cite{timmurphy.org}. However, an open issue pertains to the lack of a framework for security management in the SDN.

This article develops an intent-driven security software-defined network (IS2N) architecture by integrating blockchain and IDN. Specifically, we introduce a middle layer based on the SDN three-layer architecture and change the static connection between the controllers and switches. Also, IS2N can provide security network-level information with blockchain by invoking smart contracts and deploying consensus algorithms. Our architecture follows a closed-loop security management process from IDN, so network management is dramatically simplified. As a result, the platform brings several benefits, including security and transparency. The main contributions of this work are summarized as follows:

\begin{itemize}
    \item We develop a four-layer IS2N architecture based on the standard three-layer SDN architecture and introduce blockchain as a security component, which adds security functions such as snapshot storage and network security defense to the original SDN without degrading its performance.
    
    \item We introduce the concept of IDN into the management of IS2N, creating an intent-based security management mechanism to improve IS2N's automation.    
    
    \item We implement the IS2N four-layer architecture by deploying a blockchain-based middleware that permits transparent interactive data forwarding between the controller and switch and the submission of interactive data to the blockchain. The interactive data determine if the controller attempting network access is registered.  
    
\end{itemize}

The rest of this paper is structured as follows: Section II reviews the existing researches on blockchain and IDN. In Section III, a new IS2N architecture is presented. Section IV describes an intent-driven closed-loop security management mechanism that exists in the IS2N. The efficiency of IS2N is demonstrated in Section V through simulation results. Lastly, in Section VI and Section VII, we draw conclusions from this study.

\section{Related Work} \label{section:related}

This section presents a brief overview of current blockchain and IDN research. The first section will attempt to introduce the security functions of blockchain, while the second will focus on the IDN's essential components and applications.

\subsection{Blockchain for SDN Security}

Blockchain is a distributed ledger that stores and distributes non-duplicable data \cite{tong2020b}. Before transactions are added to the ledger, they must be requested, encrypted, and verified by other blockchain nodes. When the majority of nodes agree that a transaction is valid, a new block is added to the network and distributed. Blockchain provides efficient mechanisms for ensuring data consistency, security, and storage synchronization, in addition to innovative network security solutions. In practice, a blockchain serves the following essential functions in an SDN network:

\textbf{Certificate Authority:} Conventional certificate authorities, such as public key infrastructure (PKI), improve the administration of centralized storage, certificate issuance, and revocation. Trust issues between certificate authorities make it challenging to achieve cross-certification. Since PKI uses asymmetric encryption, theft of the private key may result in the leakage of access rights. Blockchain acts as a certificate and identity verification authority. For example, blockchain can prevent unauthorized switches from accessing the SDN, thus preventing attackers from employing illegal controls to launch distributed denial-of-service attacks on the SDN \cite{tong2020b}.
 
\textbf{Security Storage Center:} In current SDNs, network-level information is stored in controllers, while flow table information is stored in switches. At the control and data forwarding layers, there is no suitable mechanism to guarantee the integrity of flow rule information within the controller. Forwarding devices, other than the controller in the SDN, are unable to determine whether flow rules have been tampered with by attackers.  As a result, attackers can tamper with the flow rules information of a switch, potentially causing network failures. The blockchain can improve the security of data storage. For example, BigchainDB implements a database that leverages blockchain \cite{mcconaghy2016bigchaindb}. Moreover, detecting the topology information in the blockchain ensures a faithful relationship between switches and controllers. Previously, others introduced blockchain to ensure the consistency of multi-controller strategies in multi-controller scenarios, resulting in the accuracy of the issued policies \cite{azab2019towards}. In BMC-SDN, blockchain is linked to SDN controllers through ONOS application\cite{singh2020deep}.
 
A blockchain can enhance the traceability, immutability, and security of the SDN. Since blockchain is a highly secure distributed ledger, integrating SDN and blockchain can increase network security by removing interference and manipulation by third parties. This will protect network activity from malicious activity. By recording network state and transactions, the blockchain-based SDN increases network dependability. Due to the requirement for blockchain transaction confirmation, the combination of SDN and blockchain increases network complexity and network delay. As a result, we concentrate on the implementation complexity and network latency problems.

\subsection{Intent-Driven Network Management}

\begin{figure}[t]
	\centering
	\includegraphics[width=1.0\linewidth]{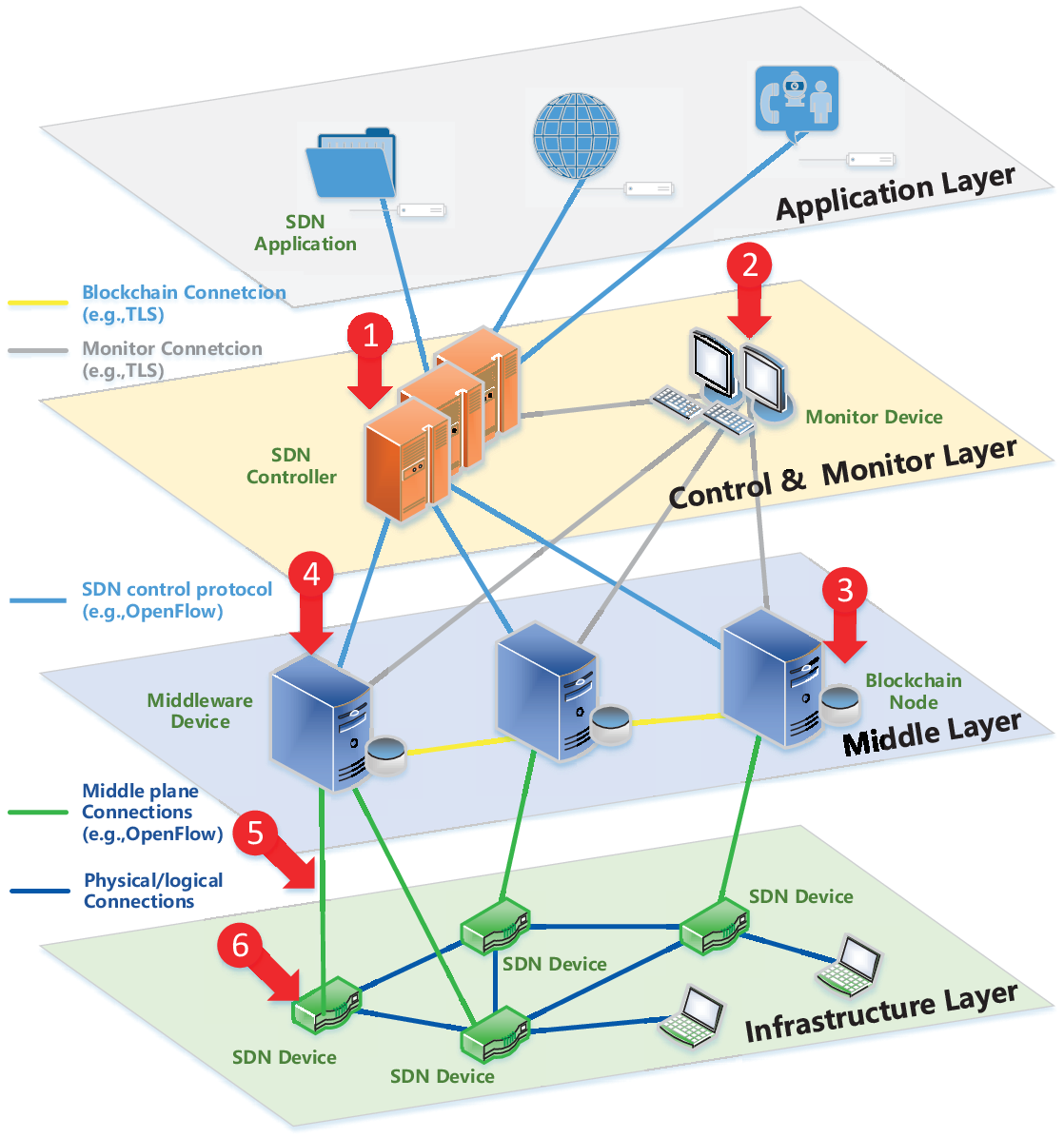}
	\caption{Logical view of the IS2N architecuture.}
	\label{arc}
\end{figure}

Although SDN empower the network programmability \cite{BlockSDN}, in practical deployment, thousands of parameters are needed to be implemented manually in network products, and configuration failures are inevitable. Future networks require minimal interruption and personnel involvement. Driven by technologies and demands, IDN changes the network architecture and affects network development quickly \cite{2020Pang}. In the IDN loop, the implementing intent requires full life-cycle verification. During this process, it may be necessary to collect network state information to identify anomalous behavior and specify corresponding policies to resist such behavior. The middleware follows the intent-driven network loop's requirements to formulate strategies that meet users' security intent and schedules network functions and implement network policies to resist attacks. According to the Gartner report, the closed-loop control methodology in IDN improves the network automation \cite{timmurphy.org}. A complete IDN system provides four main capabilities:

 \begin{itemize}
 \item  \textbf{Translation and Validation:} The IDN system takes a high-level policy as the input from the end-users and converts it to the necessary network configuration. The system then generates and validates the resulting configurations for correctness.
 
 \item \textbf{Automated Implementation:} The IDN system configures the appropriate network changes across the existing network infrastructure, typically implemented via network automation and orchestration.
 
 \item \textbf{Awareness of Network State:} The IDN system ingests real-time network status for systems under its administrative control and is protocol independent.
 
 \item \textbf{Assurance and Dynamic Optimization:} The IDN system continuously validates (in real-time) that the original business intent of a system is being met and can take corrective actions (blocking traffic, modifying network capacity, or notifying) when the intent is not met.
\end{itemize}

The IDN is an emerging concept applied and studied in different contexts, for instance, network management \cite{2020Pang}, cloud management \cite{elkhatib2017charting}, and SDN management \cite{heorhiadi2018intent}. In addition, the IDN has some initial studies in cybersecurity. Geo-Blocking proposes a system with high-level intent to protect the SDN \cite{kumar2017implementing}.

In general, combining the blockchain with the IDN is the developing trend of the SDN\cite{tong2020b}, whose advantages can be inherited simultaneously. Blockchain can enhance the security of an SDN. While the number of network devices and services increases, the management mechanism of the IDN becomes flexible and scalable, reducing the cost of configuring network settings. The IDN also assists in selecting appropriate consensus algorithms for the blockchain\cite{scheid2020controlled}.

IDN can achieve automatic network configuration and independent resource allocation to maximize network resource utilization. Since IDN rely on user data, the leakage of user data can result in severe network security issues. Therefore, it is essential to fully utilize blockchain to store intent and related data.

\section{Architecture with Middle Layer and Blockchain} \label{section:arc}

This section discusses the design of the IS2N in detail. We implement the middle layer based on three-layer SDN architecture, resulting in a four-layer IS2N architecture. Due to the middle layer, the IS2N can connect to the blockchain and store a network snapshot; the IS2N has dynamic middleware connections and can eliminate elements when they are attacked. Following that, the key capabilities of IS2N are refined and introduced.

\subsection{Architecture Design}

Herein, we construct a tailored IS2N architecture for SDN by integrating blockchain and a middle layer. In addition to the network devices in the SDN, blockchain nodes support the complete corresponding workflow. We introduce a new layer in the three-layer SDN architecture. As shown in Fig. \ref{arc}, a logical view of the IS2N includes an Application Layer (AL), a Control and Monitoring Layer (CML), a Middle Layer (ML), and an Infrastructure Layer (IL).

 \textbf{Application Layer:} The AL provides a northbound programming interface for users to express intent in text, speech, or graphics. The AL manages network elements through northbound programming interfaces and develops various business applications.

 \textbf{Control and Monitor Layer:} In the CML, the architecture's core control and monitor functions are placed in the SDN controller clusters \textcircled{1} and monitoring system clusters \textcircled{2}. Specifically, the CML monitors the network elements in real time and develops appropriate network policies. The connections \textcircled{5} between the controller clusters and the middleware devices still follow an SDN control protocol such as OpenFlow. The sFlow protocol is supported in the interface between the monitoring system and the middleware. The controller clusters are mainly responsible for network control, including formulating and distributing the network forwarding strategies. As shown in the logical view of Fig. \ref{arc}, the monitor device clusters follow an intent-driven management process, monitor the network elements, and ensure the availability of the network. In the event of attacks and intrusions, the monitor device clusters translate and refine the network-protecting intent or the controlling intent generated by the network manager. After reasoning, a configuration policy is formed, calculated, and configured into the network devices. The execution of the policy is further verified in the CML, and the performance reports are released.

 \textbf{Middle Layer}: The ML comprises middleware \textcircled{4}, which is the hub of collecting and forwarding information. The middleware can be deployed as a blockchain node \textcircled{3} or connected to the blockchain via the interface. The SDN interaction information, such as OpenFlow messages between controllers and switches, is uploaded to the blockchain by calling smart contracts. It is also responsible for registering and mapping controllers and switches. The middleware can actively disconnect the SDN control path and the middleware connection path. It is necessary to safeguard the credibility of information, such as network elements and identity information. More specifically, the ML can call smart contracts to query network security snapshots stored in the blockchain and enable network backtracking.
 
 \textbf{Infrastructure Layer}: The IL is comprised of physical devices, hosts, and physical connection links \textcircled{6}. It is mainly responsible for forwarding network traffic. The real-time status of switches is also stored on the blockchain via ML. 
 
 The inclusion of the ML serves a dual purpose. One purpose is to create an interface for SDN to communicate with the blockchain and enable on-chain storage of network multi-information; the other purpose is to split the fixed matching relationship between controllers and switches to enable the access control. 

\begin{figure}[t]
	\centering
	\includegraphics[width=1.0\linewidth]{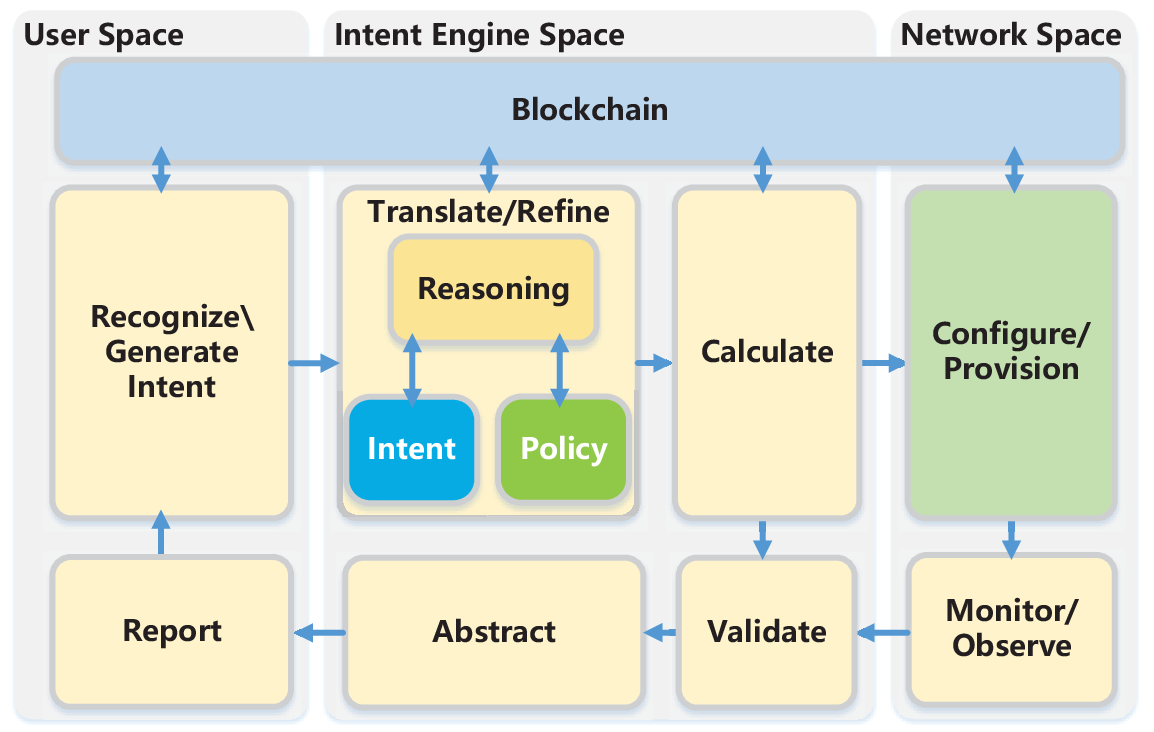}
	\caption{The view of intent-driven security management mechanism in the IS2N.}
	\label{loop}
\end{figure}

\subsection{Key Capabilities}
The IS2N enables the following key capabilities: Network element information registration and access control, network snapshot storage, and dynamic middleware connection.

\textbf{Network Element Information Registration and Access Control}: The proposed system can provide devices security registration. Each new device joining the network needs to go through the registration process. The blockchain must store each switch ID, the controller ID, and the related information. Only nodes that are registered and verified on the blockchain can participate in the standard network forwarding and flow table distribution. Simultaneously, it is necessary to analyze the message the network devices sends to determine whether the device is hijacked. The middleware can remove malicious (hijacked) nodes, thereby alleviating potential network threats. 

\textbf{Network Snapshot Storage}: The IS2N stores network security snapshots. A network security snapshot is part of the current state of the network, consisting of traffic information, interaction protocol information, and other information about the network state. The IS2N provides a more robust security solution for network security event monitoring by decoupling the southbound protocol. All control commands and network status information can be queried based on snapshots, providing more reliable data for detection algorithms and enabling more accurate attack traceability. The overall view of the SDN and the management strategies of the network operator (load balancing, routing, etc.) are also stored on the blockchain as snapshots. The snapshots can also aid in the recovery of network's state. Furthermore, we deployed different smart contracts to store multiple types of network snapshots on demand.

\textbf{Dynamic Middleware Connection}: We implemented the interaction between the blockchain and the network control system through middleware. In addition, the middleware can specify the connection between the controllers and the switches. In the conventional SDN, the mapping relationships between switches and controllers are pre-configured and cannot flexibly adapt to the network conditions. The middleware implements the controller's load balancing according to the network's specific demands and dynamically determines the link relationship between switches and controllers. Conventional controllers can reject abnormal forwarding nodes or cut-off links. However, due to the equal power of controllers, they cannot eliminate other controllers. IS2N allows SDN to reject controllers, which means cutting off the control links and solving the controller's potential hijacking.

The key capabilities of the IS2N provide novel approaches to SDN network security. The blockchain-based registration of network element information ensures the security attributes of network elements and improves the security of the SDN from the viewpoint of access. The blockchain-based storage of network information forms a snapshot to ensure that the network information is not tampered with. The dynamic middleware connection allows the IS2N to cut the controller's link actively, thereby protecting the control plane, as detailed in Section \ref{section:loop1}.

\section{Intent-Driven Security Management Mechanism Design}
\label{section:loop1}

The network security management approach of the IS2N is built on the IDN, beginning with the intent to increase automation. The IS2N management method is separated into three areas: the user, the intent engine, and the network space, as shown in Fig. \ref{loop}.

\subsection{Lifecycle of Intent in IS2N}

\begin{itemize}
    \item \textbf{Recognize and Generate Intent:} The intent can be recognized or generated by monitoring clusters in the user space. The identification of intent aims to understand the state of the network. The intent generation is based on the current network state to identify possible security events and the need of the network's intent. The generated or identified intents must be uploaded to the blockchain for subsequent validation and inspection.

    \item \textbf{Translate and Refine Intent:} Translating and refining intent are the next steps. Since intent is an abstract policy expression, it cannot be directly deployed into the network. In the intent intent translation and refinement module, intent and policy are matched. This process occurs in the intent engine space.

    \item \textbf{Configure and Provision Intent:} 
    It is necessary to calculate the policy parameters before policy deployment. For example, the link re-policy requires calculating new paths and flow rules for specific switches, and then distributing them to the corresponding switches in the IL. The process of configuring networks takes place in the network space.

    \item  \textbf{Monitor and Validate Intent:} The CML senses changes in the network and verifies the validity of the intent. Security managers will refer to the reports generated by the CML. Additionally, to maintain the current intent, the CML adjusts the results of the parameter calculation according to the dynamic changes in the network. The CML will always follow the present intent to protect the network until the arrival of new intents. Then, it moves to the following intent management loop.
\end{itemize}

\subsection{Component of IS2N}

The IS2N platform comprises critical components, including one intent engine, one controller with monitor, one blockchain, middleware, and two middleware interfaces. This section mainly introduces the interaction logic between the functions of each component.

 \begin{figure}[t]
	\centering
	\includegraphics[width=0.9\linewidth]{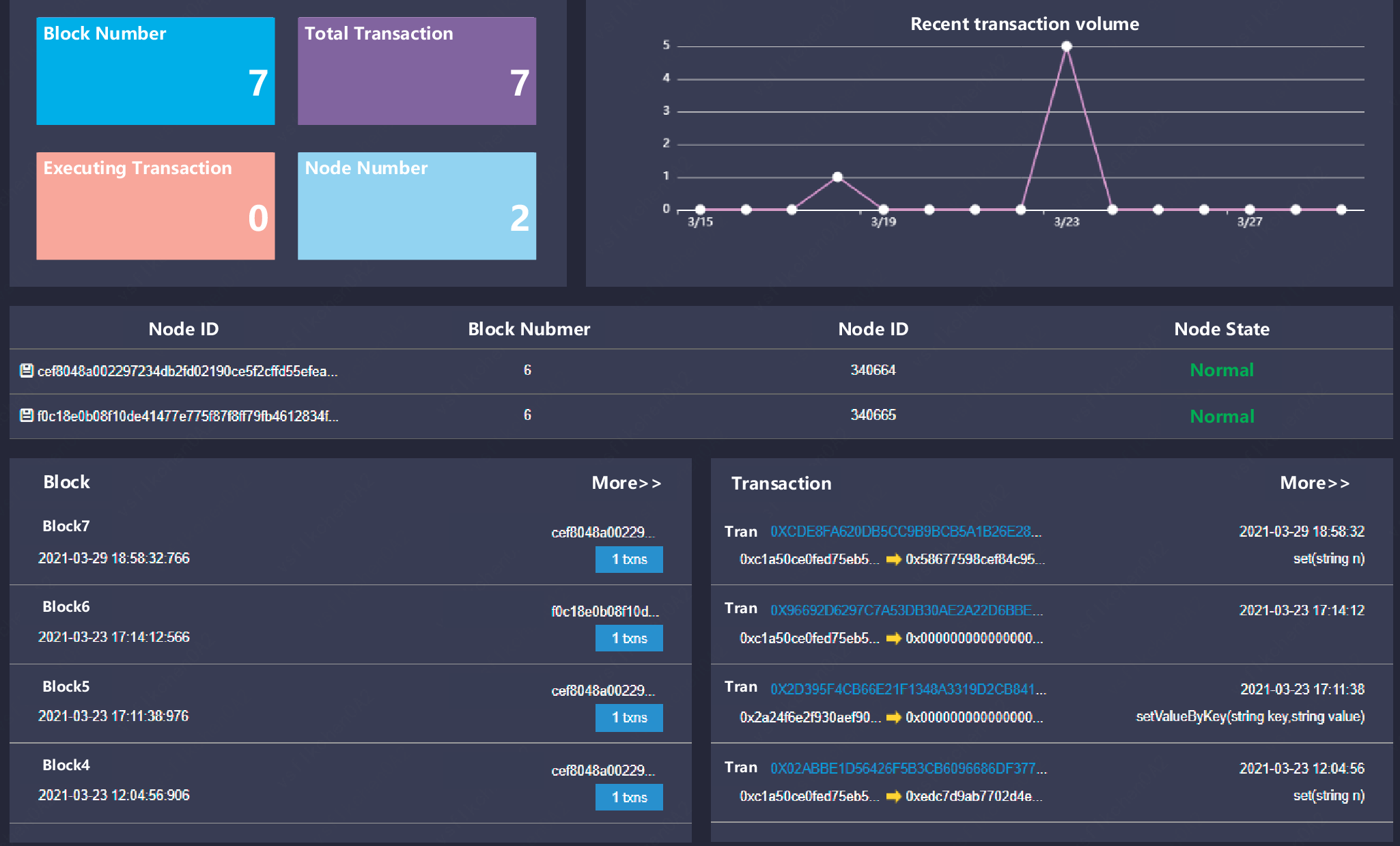}
	\caption{Blockchain information displayed on the IS2N graphical user interface.}
	\label{fig:gui}
\end{figure}

\begin{itemize}

    \item \textit{The Graphical User Interface (GUI)} primarily presents the system's operation in a simple manner. The number of consensus nodes, total number of transactions, current block number, and the number of block transactions on the blockchain can all be found in the GUI, as shown in Fig. \ref{fig:gui}. Our platform has successfully saved a few network snapshots, proving that blockchain is practical. Managers can look at a specific network snapshot by looking up the hash value.
   
    \item \textit{Intent Engine} is responsible for reasoning intents and allows users to express the goals related to their intents. The user can fill in abstract high-level security requirements, such as removing device or recalculating paths.

    \item \textit{Controller and Monitor} are responsible for receiving and monitoring intents, calculating the most secure policy, and developing configuration scripts based on the real-time state of the network. In the intent misconfiguration event or unanticipated network security events, the intent can be automatically adjusted to ensure its correct execution without the intervention of an administrator. 

    \item \textit{Blockchain} stores any changes of the network state that occurs during the life-cycle of intents, including intent, policy, and flow table modification. The flow table will be uploaded to the blockchain to ensure the policy's consistency. The number of switches and the device information are used as the block-header, and the switch's port information, transmission rate, and link status information are transmitted to the blockchain as block-body. The security of IS2N is ensured through the blockchain's identity verification and policy consistency.

    \item \textit{Middleware} implements the middle layer and emulates the functions of controllers and switches. Therefore, it is a transparent device, and controllers and switches can not feel the presence of the middleware during the connection process. The middleware maintains the following two interfaces during runtime.

\end{itemize}

 \textbf{The interface between controller and middleware}: The controller sends the registration information to the middleware. After receiving the registration information from the controller, the agent middleware sends the information to the blockchain, which invokes a smart contract to securely store the relevant information.

 \textbf{The interface between switch and middleware}: The middleware initiates a connection process with the switch. Next, the middleware simulates the controller to complete the handshake process with the switch, sending the pertinent information to the blockchain. The middleware additionally records the switch data and the simulated handshake data at this time in the blockchain.

The manager submits intents to the intent engine. Subsequently, the policy is forwarded to the controller after reasoning by the intent engine. While the controller receives a successful storage receipt from the blockchain, the configuration file is sent to the underlying device. The middleware collects the network state data and uploads it to the blockchain, invoking the contract and waiting for the receipt. The monitoring system analyzes the data and verifies if the intent needs to be corrected.

\section{Intent-Driven Security Software-Define Network Implementation} \label{section:implement}

To demonstrate the capabilities of the IS2N architecture, we develop a middleware, deploy a blockchain, and realize the interface between them. We compare the consensus times of two blockchain algorithm to store OpenFlow messages as network snapshots. Then, we simulate DDoS attacks and sever the connection to the targeted network components by middleware.

\subsection{Platform Implementation}

To evaluate the performance of the proposed IS2N, we built the simulation platform on ONOS controller cluster. We set up 6 Open vSwitches, 25 hosts in Mininet, and 7 to 31 blockchain nodes, where the blockchain nodes are deployed as Docker containers based on the open-source project FISCO-BCOS \cite{Fisco.org}. We select ONOS as the platform's controller because ONOS is more effective to ODL in terms of fault tolerance and delay\cite{2018fault}.

 \begin{figure*}
 	\centering
 	\includegraphics[width=0.9\linewidth]{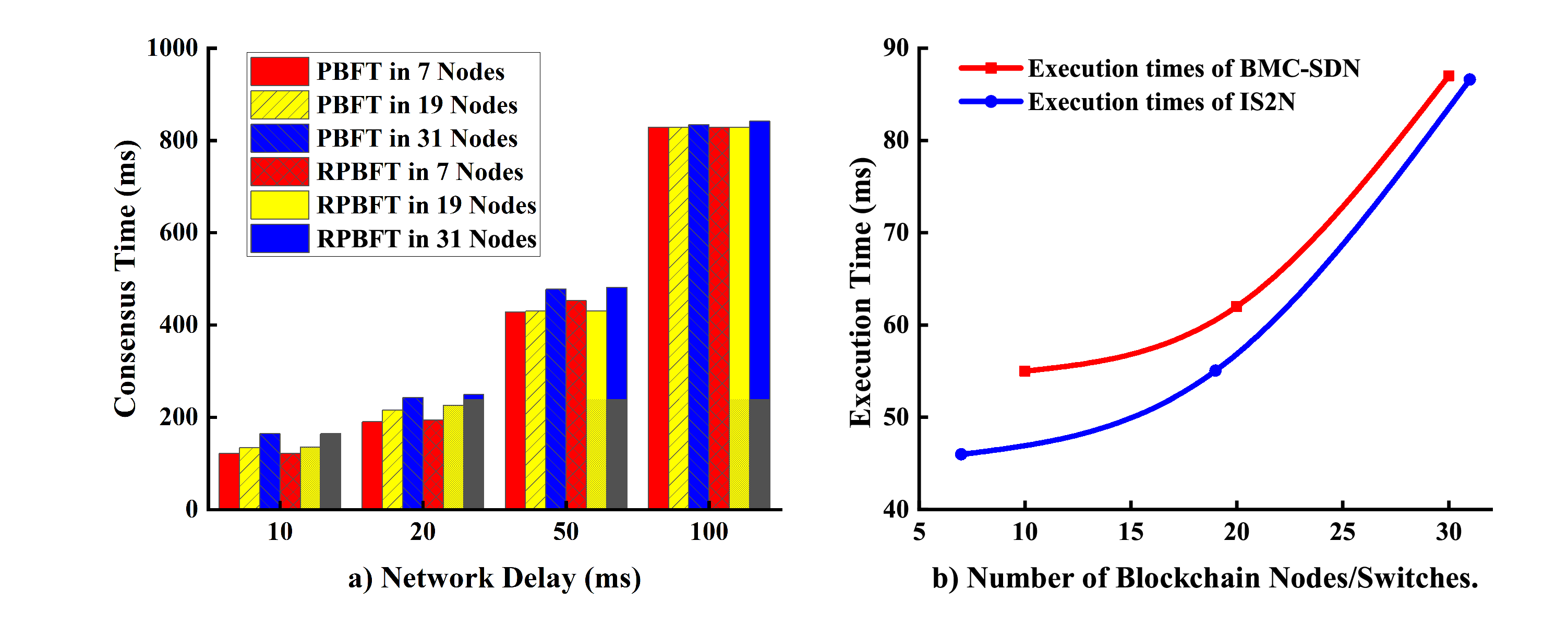}
 	\caption{Comparison of the delay performance of IS2N with different consensus nodes and OpenFLow message with PBFT: a) Comparison of different consensus times algorithms used in IS2N; b) Comparison of execution times between BMC-SDN and IS2N.}
 	 \label{fig:fig4-a-b}
 \end{figure*}
 
\subsection{Performance Evaluation}

\subsubsection{The Performance of Consensus Time} 

The consistency algorithms is the foundation of IS2N platform. Therefore, we test the effect of different consistency algorithms on the IS2N platform to demonstrate the performance with a set of preliminary results. First, we apply the blockchain with two consensus algorithms: practical Byzantine fault tolerance (PBFT) and redundant practical Byzantine fault tolerance (RPBFT). We deploy 7, 19, and 31 blockchain nodes to satisfy the requirement of $3N+1$. The transmission delay between blockchain nodes are set to 10, 20, 50, or 100 milliseconds, corresponding to the impact of packet loss on the network condition: little to no impact, a negligible effect on the network, a significant effect on the network, and an enormous effect on the network. In each simulation, multiple reading and writing operations were executed, and their respective delays were recorded. As shown in Fig. \ref{fig:fig4-a-b} (a), the consensus time is positively correlated not only with the number of consensus nodes but also with the network transmission delay. Specifically, the performance of PBFT is better when the number of nodes is small. As the number of nodes increases, the gap between the two algorithms gradually becomes narrower. Therefore, the platform with the PBFT consensus algorithm is more suitable for scenarios with a lower number of nodes.

We compared our architecture with BMC-SDN, another blockchain-based SDN architecture\cite{singh2020deep}. In contrast, IS2N introduces middleware between controllers and switches to carry out blockchain functions. Our architecture allows for interrupting the connection between controllers and switches through middleware deployment, as well as for actively changing controllers. Although introducing blockchain increases network transmission delay, our performance is comparable to BMC-SDN architecture, as shown in Fig. \ref{fig:fig4-a-b} (b). However, we focus on addressing security issues affecting the network's control plane. For example, middleware can interrupt the link to prevent network from damage in the controller attacked event.

\subsubsection{The Performance of Network Element Information Registration and Snapshot} 

To validate the information registration performance of the network element, we measure the write and read delay of various messages on the platform. The experimental procedure includes deploying the blockchain system and smart contract, executing the middleware and ONOS controllers, invoking the relevant methods of the smart contract through the Go program, and recording the delay.

\begin{figure}
	\centering
	\includegraphics[width=1.0\linewidth]{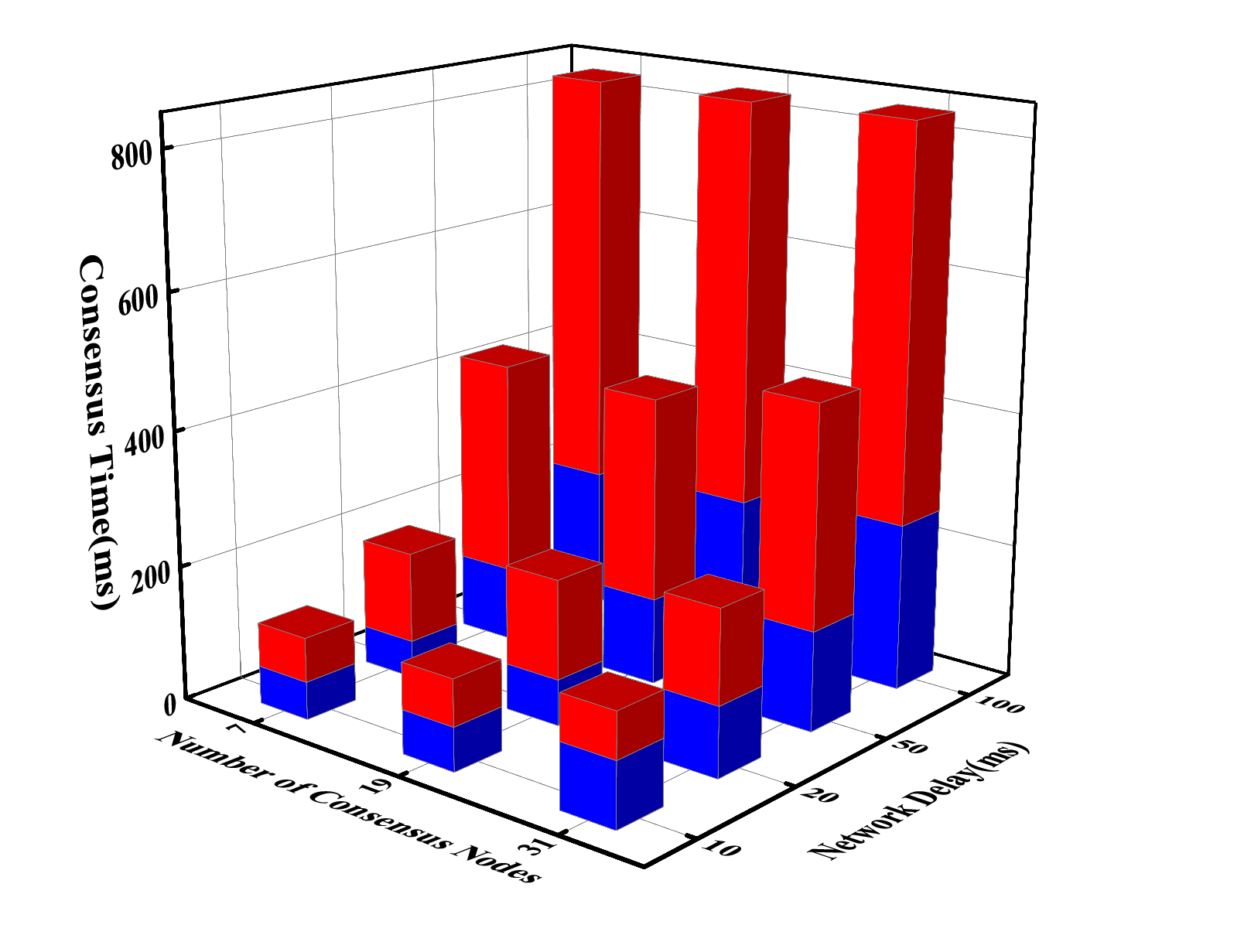}
	\caption{Platform consensus time when storing typical OpenFlow message.}
	\label{fig:fig5}
 \end{figure}

Due to the better consensus time of the PBFT at the small number of nodes, we further deploy PBFT on the IS2N platform. The interaction between controllers and switches is delay-sensitive in the SDN, and introducing middleware may affect the interaction time of the OpenFlow protocol. Therefore, we simulate the delay performance with different numbers of block nodes. As depicted in Fig. \ref{fig:fig5}, the delay remains in the millisecond range. The blue histograms indicate the amount of time required to upload metadata and establish consensus. The red histograms reflect the additional time required to create snapshots and register network elements, such as controllers and switches via OpenFlow messages. The additional delay is the cost of implementing the middleware. The consensus time increases with the increase of nodes and the network delay since the consensus algorithm needs to reach consensus in at least $1/3$ of the nodes, and the network delay is a bottleneck to consensus.

\begin{figure*}
	\centering
	\includegraphics[width=0.75\linewidth]{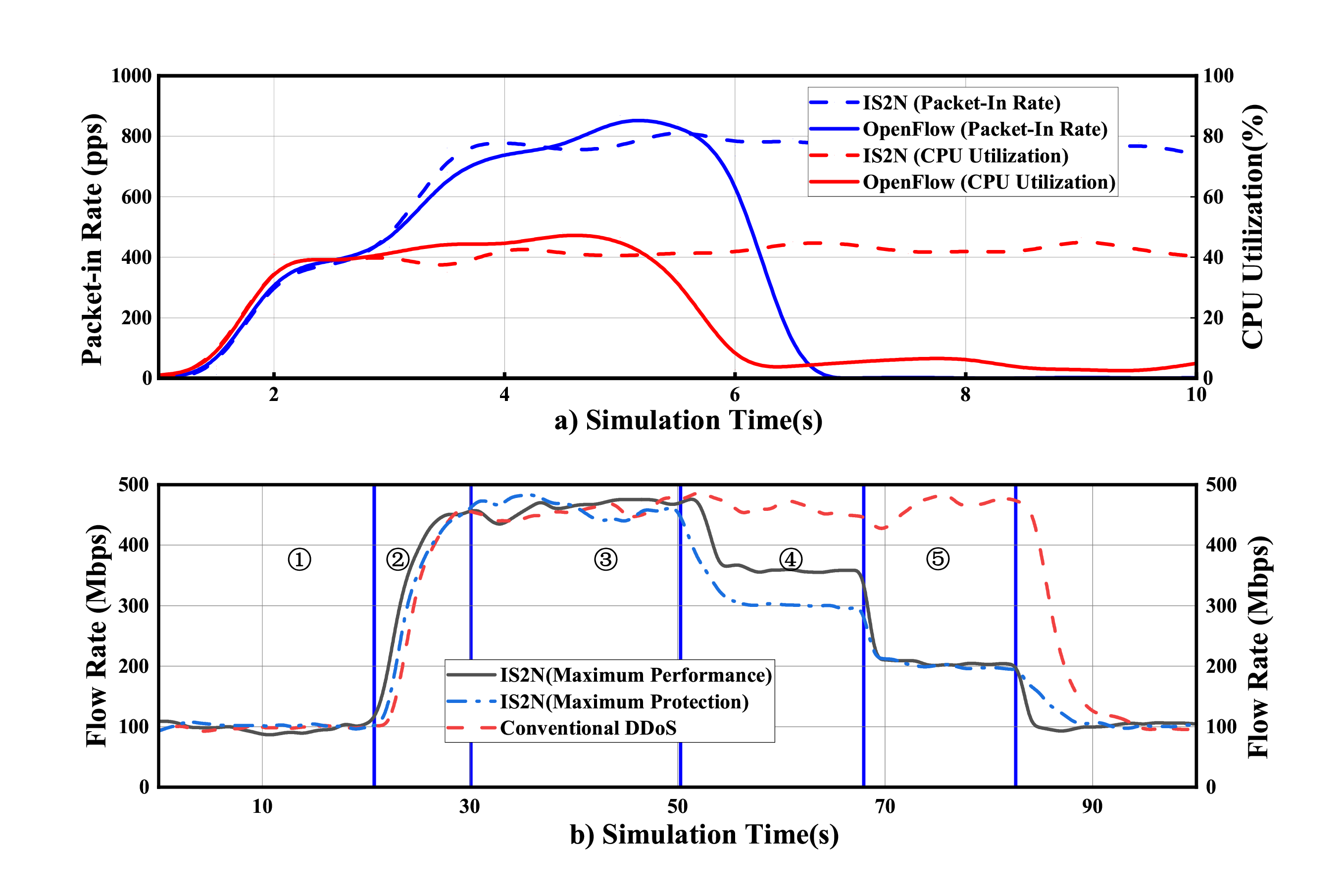}
	\caption{Simulation results security performance and policy flexibility of the IS2N: a) The comparison between OpenFlow and IS2N in packet rate and CPU utilization after DDoS attack;  b) Simulation results using IS2N policies: traffic volume measured in a specific link leading to the victim IP address.}
	\label{fig:fig-6}
\end{figure*}

Likewise, an increases in the number of nodes results in an increase in the consensus time. This is illustrated in Fig. \ref{fig:fig-6} (a), where we store a snapshot containing OpenFlow messages. The length of OpenFlow messages and the processing delay of switches and controllers result in the increased interaction time. By default, the interval for the switch or the controller to send OpenFlow Echo-Request packets is five seconds. If there is no response in three Echo-Request packets, the link is considered disconnected. Consequently, our platform can be used as an overlay technology for SDN and does not impact the interaction process between the controller and the switch.

\subsubsection{Security Performance}

IS2N is primarily concerned with control plane security, so we validated it by simulating DDoS attacks against the control plane. We simulate a DDoS attack in which multiple forged source addresses are sent to victim hosts located on separate switches. We targeted the control plane by generating an OpenFlow new packet $packet\_in$, rather than simply forging data packets. The unknown packets will be routed to the controllers, thereby consuming controller resources. 

We compare the IS2N with the OpenFlow mechanism under DDoS. The CPU utilization and received $packet\_in$ rate of the controller are depicted in Fig. \ref{fig:fig-6} (a). We launch a DDoS attack from two seconds to the end. It can be clearly seen that the $packet\_in$ rate and CPU utilization keep consistent at a high level after the attack. In contrast, these two metrics decrease rapidly after five seconds because the defense $flow\_mod$ have been installed on the switch via middleware. This can be validated by $ovs-ofctl dump-flows bridge$ command on the OpenvSwitch. The defense flow table entry consists detected inport. As described in Section \ref{section:arc}, due to the dynamic middleware connection, middleware can successfully stop DDoS attack on the control plane. Therefore, the IS2N effectively mitigates DDoS attacks on the control plane.

\subsubsection{Policy Flexibility}
In addition to effectively resisting DDoS attacks on the control plane, IS2N can select various policies to prevent attacks flexibly. In order to evaluate various limiting policies, two scenarios were considered. Both policies aim to quickly restrict abnormal network traffic to acceptable levels. The network is stable \textcircled{1} at the beginning and we launch a flooding attack \textcircled{2}. The defense mechanism of IS2N is implemented, and then the abnormal link is first limited \textcircled{3}, then the abnormal IP address is found \textcircled{4}, and the abnormal traffic is further idle. Finally, the abnormal flow is accurately identified \textcircled{5}. \textcolor{black}{We used different preferences for the IS2N middleware in each scenario that included the controller capability.} In the $Maximum {\rm{ }} Performance$ scenario, the middleware had a stronger preference for the maximum protection soft goal. As a consequence, the plan with the highest utility is high, as can be seen at around 55s in Fig. \ref{fig:fig-6} (b). The execution of this plan was enough to achieve the goal of restricting the access to the victim IP address. In the $Maximum {\rm{ }} Protection$ scenario, we set a stronger preference for the maximum performance soft-goal, as can be noted at around 55s in Fig. \ref{fig:fig-6} (b). The detect action was executed within this plan, and detected that the victim IP address remains an outlier. Later, additional deep packet detection functions and attack source tracing processes will be initiated, enabling specific attack IP addresses and abnormal flow to be identified, thus reducing the abnormal traffic gradually until the network returns to a stable state. Compared to conventional DDoS defense, IS2N can shorten the time that the network is in an unacceptable state, and the strength of the defense policy can be adjusted according to the requirements.

\section{Challenges and Future Work}

This article explores the combination of IDN and blockchain to enhance SDN security. Nonetheless, several open issues still require attention for future development. The primary focus should be on enhancing the security level of the physical nodes where the blockchain is located, which can help overcome the bottleneck of its performance in SDN. Afterwards, smart contracts should be manually written and deployed onto the blockchain in advance. Future studies could explore the automation of smart contract programming and deployment. The perspective of automatic code generation and formal verification techniques should be considered in such studies. In addition to storing and synchronizing information, the blockchain can deploy more functions. Additional studies are necessary to enable learning algorithms on the blockchain to detect abnormal behaviors of controllers and switches. In our future work, we plan to consider other consensus algorithms and the blockchain-based SDN and blockchain-based blockchain will be investigated in greater detail.

\section{Conclusion}\label{section 7}

This article discussed the feasibility of combining blockchain and intent-driven networks into software-defined networks. We proposed a four-layer architecture with a blockchain-based middle layer. 
The intent-driven mechanism was used to operate the security software-defined networks. By introducing a blockchain, we separated a constant mapping between the controllers and the switches, which allowing for dynamic middleware connections, multiple information storage snapshots, and synchronization issues. The performance of consensus algorithms for the proposed platform suggested that security middleware have little effect on software-defined networks. Moreover, middleware could promptly disable attacked network components to safeguard controller resources and enable control capabilities. Therefore, developing the blockchain-based mechanisms and intent-driven network improved the security of software-defined networks.

\bibliographystyle{IEEEtran}
\bibliography{IEEENetwork-IS2N-YANBO-FIN}

\end{document}